\documentclass[12pt]{article}
\def\poinc{Poincar\'{e} }
\def\bfq {{\bf q}}
\def\bfK{{\bf K}}

\def\bfk{{\bf k}}
\def\bfp{{\bf p}}  
\def\be{\begin{equation}}
 \def \ee{\end{equation}}
\def\bea{\begin{eqnarray}}
  \def\eea{\end{eqnarray}}

\def\up{{\uparrow}}
\def\down{{\downarrow}}

\usepackage{amsfonts}
\usepackage{amssymb}

\title{\begin{flushright}{\normalsize NT@UW-02-034}\end{flushright}
Relativity, chiral symmetry, and the nucleon electromagnetic form factors}
\author{Gerald A. Miller
\\ University of Washington
  Seattle, WA 98195-1560}

\begin{document}

\maketitle
\begin{abstract}
A relativistic
  interpretation for why the proton's
  $G_E/G_M$ falls and 
$QF_2/F_1$ is approximately constant is presented. 
  Reproducing 
the observed  $G_E^n$  mandates  the inclusion of the
effects of the pion cloud. The full relativistic model with a pion cloud
provides a good reproduction 
of all of the nucleon electromagnetic form factors.
\end{abstract}

\section{Introduction}
An alternate title could be ``Surprises in the Proton''. This talk owes its
existence to the  precise, stunning and exciting recent
experimental work on measuring $G_E/G_M$ (or $QF_2/F_1$) for the proton and
$G_E,G_M$ for the neutron. My goal here is to interpret the data. Symmetries
including
Poincar\'{e} invariance and chiral symmetry will be the principal tool
I'll use. This talk is based on  three papers
\cite{Frank:1995pv},\cite{Miller:2002qb},\cite{Miller:2002ig}.

If, a few years ago,
one had asked participants at a meeting like this about the $Q^2$ dependence
of the proton's $G_E/G_M$ or $QF_2/F_1$. Almost everyone one have answered that
for large enough values of $Q^2$, $G_E/G_M$ would be flat and
$QF_2/F_1$ would fall with increasing  $Q^2$. The reason for the latter fall
being conservation of hadron helicity. 
Indeed, the shapes of the curves have been
obtained in the new measurements, except for the mis-labeling of the ordinate
axes. The expected flatness of $G_E/G_M$ holds for $QF_2/F_1$, and the
 quantity $G_E/G_M$ falls rapidly and linearly with $Q^2$. This
 behavior needs to be understood!

\section{Outline}
\label{outline}
I will begin with a brief introduction to  Light Front Physics.
Then I will discuss a particular relativistic model of the nucleon,
and  proceed to apply it to the proton form factors, with the aim of
providing a qualitative understanding of the salient experimental features.
The same model fails to reproduce the neutron $G_E$ unless the effects of the 
 pion cloud are included. The combination 
of relativistic effects with those of the pion cloud lead to a model that is
able to reproduce all of the nucleon electromagnetic form factors.
\section{Light Front }

Light-front dynamics
is a relativistic many-body dynamics in which fields are quantized at a
``time''=$\tau=x^0+x^3\equiv x^+$. The $\tau$-development operator
 is then given by
$P^0-P^3\equiv P^-$. These equations show the  notation that    
a four-vector $A^\mu$ is expressed  as
$ A^\pm\equiv A^0\pm A^3.$
One quantizes at $x^+=0$ which is a light-front, hence the name ``light front
dynamics''.
 The canonical spatial variable must be orthogonal to the time variable,
 and this is given by 
$x^-=x^0-x^3$. The canonical momentum is then $P^+=P^0+P^3$. The other
coordinates are  ${\bf x}_\perp$ and  ${\bf P}_\perp$.

 The most important  consequence of this is that the
 relation between energy and
momentum of a free particle is given by:
$ p_\mu p^\mu=m^2=p^+p^--p_\perp^2\to  p^-={p_\perp^2+m^2\over p^+},$
 a relativistic kinetic energy which does not contain
a square root operator. This
allows the  separation of  center of mass
and relative coordinates, so that the computed wave functions are frame
independent.

The use of the light front is particularly relevant for calculating form
factors, which are probability amplitudes for an nucleon to absorb a four
momentum $q$ and remain  a nucleon. The initial and final nucleons 
have different total momenta. This means that the final nucleon is a
boosted nucleon, with 
 different wave function than the initial nucleon.  In general, performing
 the boost is difficult for large values of $Q^2=-q^2$. However the light
 front technique allows one to set up the calculation so that the boosts are
 independent of interactions. Indeed, the wave functions are functions of
 relative variables and are independent of frame. 

\section{Definitions}

\newcommand{\boldsigma}{\mbox{\boldmath $\sigma$}}

Let us define the basic quantities concerning us here. These are the
independent form factors defined by
\begin{equation}
\left< N,\lambda ' p' \left| J^\mu \right| N,\lambda p\right> =
\bar u_{\lambda '}(p') \left[ F_1(Q^2)\gamma^\mu + {\kappa F_2(Q^2) \over
2 M_N}i\sigma^{\mu\nu}(p'-p)_\nu \right] u_\lambda (p).
\ee
The 
Sachs form factors are defined by the equations:
\bea
G_E = F_1 - {Q^2 \over 4M_N^2}\kappa F_2,\; 
G_M = F_1 +  \kappa F_2\label{sachsdefs}.\eea

There is an alternate 
  light front interpretation, based on field theory, in which one uses
the  ``good" component  of the current, $J^+,$ to 
 suppress  the effects of quark-pair terms. Then, using nucleon light-cone
 spinors: 
\begin{eqnarray}
F_1(Q^2) ={1 \over 2P^+}\langle N,\uparrow\left| J^+\right| N,
\uparrow\rangle, \quad
Q\kappa  F_2(Q^2) ={-2M_N \over 2P^+}\langle N,\uparrow\left|
J^+\right| N,\downarrow\rangle.
\end{eqnarray}
The form factor $F_1$ is obtained from the non-spin flip matrix element, while
$F_2$ is obtained from the spin-flip term.

\section { Three-Body Variables and Boost}
We use 
light front coordinates for the momentum 
of each of the $i$ quarks, such that
${\bf p}_i =
(p^+_i,{\bf p}_{i\perp}), p^-=(p_\perp^2+m^2)/p^+.$ The  total
(perp)-momentum
is 
$\bf {P}= {\bf p}_1+{\bf p}_2+ {\bf p}_3,$ the plus components of the
momenta are denoted as 
\be \xi={p_1^+\over p_1^++p_2^+}\;,
\qquad
\eta={p_1^++p_2^+\over P^+},\ee
and the perpendicular relative coordinates are given by
\be {\bf k}_\perp =(1-\xi){\bf p}_{1\perp}-\xi {\bf p}_{2\perp}\;, \quad
{\bf K}_\perp =(1-\eta)({\bf p}_{1\perp}
+{\bf p}_{2\perp})-\eta {\bf p}_{3\perp}.\ee
In 
the  center of mass  frame we find:
\be
{\bf p}_{1\perp}={\bf k}_\perp+\xi {\bf K}_\perp,\;\;\quad
{\bf p}_{2\perp}=-{\bf k}_\perp+(1-\xi){\bf K}_\perp\;,\quad 
{\bf p}_{3\perp}=-{\bf K}_\perp .\ee
The coordinates $\xi,\eta, \bfk,\bfK$ are
all relative coordinates so that one obtains a 
 frame independent wave function
   $\Psi({\bf k}_\perp,{\bf K}_\perp,\xi,\eta).$

Now consider the computation of a form factor, taking quark 3 to be the one
struck by the photon. One works in a special set of frames with $q^+=0$ and
$Q^2=\bfq_\perp^2$,
so that the value of $1-\eta$ is not changed by the photon. The
coordinate $\bfp_{3\perp}$ is changed to $\bfp_{3\perp}+\bfq_\perp,$
so  only one relative momentum, $\bfK_\perp$ is changed:
\bea{\bfK'}_\perp =(1-\eta)({\bf p}_{1\perp}
+{\bf p}_{2\perp})-\eta ({\bf p}_{3\perp}+{\bf q}_\perp)\;
=\bfK_\perp-\eta\bfq_{\perp},\quad
\bfk'_\perp=\bfk_\perp,\qquad  \eea

The arguments of the spatial wave function are
taken as the mass-squared operator for a non-interacting system:
\be M_0^2\equiv\sum_{i=1,3} p^-_i\;P^+-P_\perp^2=
{K_\perp^2\over \eta(1-\eta)}+
{k_\perp^2+m^2\over \eta\xi(1-\xi)} +{m^2\over 1-\eta}. \ee
This  is a relativistic version of the  square of  the center-of-mass 
kinetic energy, expressed in terms of light-front variables.
 Note that      the absorption of a photon 
changes the  value to:
\be{{M_0}'}^2=
{(K_\perp-\eta q_\perp)^2\over \eta(1-\eta)}+
{k_\perp^2+m^2\over \eta\xi(1-\xi)} +{m^2\over 1-\eta}.\ee

\section { Wave function}
Our wave function is based on symmetries. The wave function is anti-symmetric,
a function of relative momenta, independent of reference frame, an eigenstate
of the spin operator and rotationally invariant (in a specific well-defined
sense). The use of symmetries is manifested in the construction of such 
wave functions, as originally described by
 Terent'ev \cite{bere76},               Coester\cite{chun91} and their
 collaborators. 
A schematic  form of the wave function is
 \be
\Psi(p_i)=\Phi(M_0^2)
u(p_1) u(p_2) u(p_3)\psi(p_1,p_2,p_3),\quad p_i=\bfp_i  s_i,\tau_i\ee
where $\psi$ is a spin-isospin color amplitude factor,
the $p_i$ are expressed in terms of relative coordinates
(for example, ${\bf p}_{3\perp}=-{\bf K}_\perp$),  the
$u(p_i)$ are ordinary 
  Dirac spinors and $\Phi$ is a spatial wave function. The ordinary
Dirac spinors depend on the third or $z$ component of the momenta and
this is given, for example, by $p_{3z}=
{1\over2}[M_0(1-\eta)-{m^2+K_\perp^2\over(1-\eta)M_0}]$.

We take the the spatial wave function from  Schlumpf\cite{Schlumpf:ce}: \bea
\Phi(M_0)={N\over (M^2_0+\beta^2)^{\gamma}}\;,  
\beta =0.607\;{\rm GeV}, \; \gamma=3.5,\; m = 0.267\; {\rm GeV}.
\label{params}\eea
 The value
of 
$\gamma$ is chosen that $ Q^4G_M(Q^2) $   is approximately  constant for
$Q^2>4\; {\rm GeV}^2$ in accord with experimental data. The parameter 
$\beta$ helps govern the values of the perp-momenta allowed by the
wave function $\Phi$ and is closely related to the  rms charge radius,
and $m$ is mainly determined by the magnetic moment of the proton.

At this point the wave function and the calculation  are 
completely defined. One could evaluate
the form factors as $\langle \Psi\vert J^+\vert \Psi\rangle$ and obtain the
results. Actually the first results were obtained in 1995\cite{Frank:1995pv},
and are given as Figs.~10,11 in that reference. It is clear that the model
predicted that $G_E$ falls off much more rapidly with increasing $Q^2$ than
does $G_M$.  The result that $Q^4G_M$ is flat at high $Q^2$ is a 
consequence of using Schlumpf's wave function, 
but our values of $G_E$ for large $Q^2$ were a prediction.

So we had this  early prediction, which was
 confirmed by experimental measurements. But we didn't present an
explanation of the numerical results. This is the next task.
\section { Simplify  Calculation- Light Cone Spinors}

The evaluation  of the operator $J^+\sim \gamma^+$
 is simplified by using light cone
spinors. These      solutions of the free Dirac equation, related to ordinary
Dirac spinors by a unitary transformation,   conveniently satisfy:
\bea \bar u_L(p^+,\bfp',\lambda')\gamma^+u_L(p^+,\bfp ,\lambda)=
2\delta_{\lambda\lambda'}p^+. \eea 
To take advantage of this,
re-express the wave function in terms of light-front spinors using
the  completeness relation: 
$1= \sum_\lambda u_L(p,\lambda)\bar u_L(p,\lambda).$ We then find
  \bea
&&\Psi(p_i)=u_L(p_1,\lambda_1) u_L(p_2,\lambda_2) u_L(p_3,\lambda_3)
\psi_L(p_i,\lambda_i),\\
&&\psi_L(p_i,\lambda_i)\equiv
[\bar u_L(\bfp_1,\lambda_1)u(\bfp_1, s_1)]
[\bar u_L(\bfp_2,\lambda_2)u(\bfp_2, s_2)]\nonumber\\ \times
&&[\bar u_L(\bfp_3,\lambda_3)u(\bfp_3, s_3)]\;
\psi(p_1,p_2,p_3).\eea This is
the  very same  $\Psi$ as before, it is just that now it is
easy  to compute the matrix elements of the $\gamma^+$ operator. 

The  unitary transformation is  also known as the Melosh rotation.
The basic point is that one may evaluate the coefficients in terms of Pauli
spinors: $\vert \lambda_i\rangle,\vert s_i\rangle,$
with $\langle \lambda_i\vert { R}_M^\dagger(\bfp_i)\vert s_i\rangle
\equiv \bar u_L(\bfp_i,\lambda_i)u(\bfp_i, s_i)$. It is easy to show that
  \be \langle \lambda_3\vert 
{ R}_M^\dagger(\bfp_3)\vert s _3\rangle= \langle \lambda_3\vert
\left[ {m+(1-\eta)M_0+i{\boldsigma}\cdot({\bf n}\times {\bf p}_3)\over
\sqrt{(m+(1-\eta)M_0)^2+p_{3\perp}^2}}\right]\vert s_3\rangle.
\label{melosh}\ee
The important effect resides in the term $({\bf n}\times {\bf p}_3)$ which
originates from the lower components of the Dirac spinors. This  large
  relativistic spin effect can be summarized:
  the effects of relativity are to replace 
 Pauli spinors by     Melosh rotation operators acting on Pauli spinors.
 Thus 
\be \vert\uparrow \bfp_i\rangle\equiv{ R}_M^\dagger(\bfp_i)
\pmatrix{1\cr 0},\;\vert\up\bfp_3\rangle\ne \vert\up\rangle.\ee
In the non-relativistic limit, the Melosh rotation matrices become unit
operators and one recovers the familiar $SU(6)$ quark model.
 
\section{  Proton $F_1,F_2$-Analytic Insight}

The analytic insight is based on Eq.~(\ref{melosh}). Consider
high momentum transfer such that
$Q=\sqrt{\bfq_\perp^2}\gg \beta=560$ MeV.  {\em Each} of the quantities: 
$M_0\;,M_0'\;,\bfp_{3\perp},\;\bfp_{3\perp}$ can be of order $q_\perp,$ 
so   the spin-flip term is as large as the non-spin flip term. In particular,
($s_3=+1/2$)     
 may correspond to ($\lambda_3=-1/2)$, so the spin of the  
  struck quark $\ne$ proton  spin.
This means that
there is no hadron helicity selection rule\cite{ralston,Braun:2001tj}.

The effects of the 
  lower components of  Dirac spinors, which cause the
  spin flip term $\boldsigma\times \bfp_3$,
  are the same as having a  non-zero $L_z$,
  if the wave functions are expressed in the light-front basis.

We may now  qualitatively understand the numerical results, since 
\begin{eqnarray}
F_1(Q^2) &=& 
\int\! {d^2\!q_\perp d\xi\over \xi(1-\xi)}{ d^2K_\perp d\eta\over\eta(1-\eta)}\;
\cdots
\;\langle\up\bfp_3'\vert
\up\bfp_3\rangle \\
Q\kappa F_2(Q^2) &=& 2M_N
\int\! {d^2\!q_\perp d\xi\over \xi(1-\xi)} 
{d^2K_\perp d\eta\over\eta(1-\eta)}\;
\cdots
\langle\up\bfp_3'\vert
\down\bfp_3\rangle, 
\end{eqnarray}
where the $\cdots$ represents common factors. The term
$F_1\sim\langle\up\bfp_3'\vert\up\bfp_3\rangle$ is a spin-non-flip
term and 
$QF_2\sim\langle\up\bfp_3'\vert\down\bfp_3\rangle$ 
depends on the spin-flip term.
In doing the integral each of the momenta, and $M_0,M_0'$ can take the
large value $Q$ for some regions of the integration. Thus in the integral
\bea
\langle\up\bfp_3'\vert\up\bfp_3\rangle
\sim {Q\over Q},\;\quad
\langle\up\bfp_3'\vert\down\bfp_3\rangle\sim {Q\over Q},\eea
so that 
 $F_1$ and $ QF_2$ have the same $Q^2$ dependence. This
is shown in Fig.~1. Indeed, for $Q^2$ greater than about 2 GeV$^2$, the ratio
$QF_2\over F_1$ varies very little with increasing $Q^2$.
\begin{figure}
\unitlength1cm 
\begin{picture}(10,8)(0,-8.)
  \includegraphics{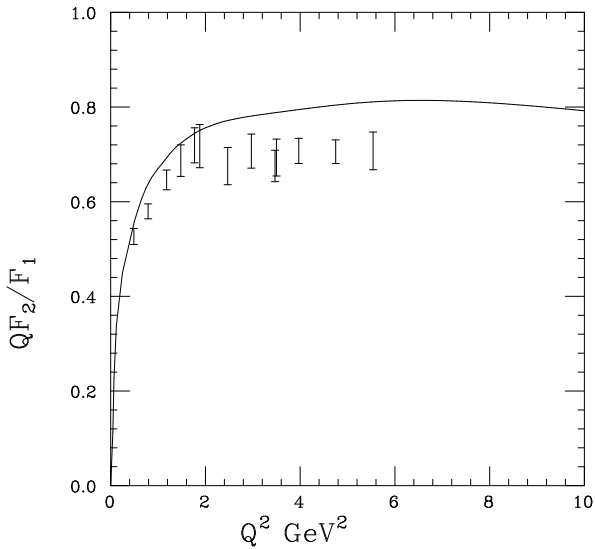}
\end{picture}
\label{fig:1}
\caption{Calculation of Refs.~$[1,2]$, 
  data 
  are from  Jones {\em et al.}$[9]$ for 
$2\le Q^2\le 3.5$ GeV$^2$
and from  Gayou {\em et al.}$[10]$. 
for $3.5\le Q^2\le 5.5\; {\rm GeV}^2$.} 
\end{figure}

\section{Neutron Charge Form Factor}
The neutron has no charge, $ G_{En}(Q^2=0)=0$, and the square of its charge
radius is determined from the low $Q^2$ limit as
$G_{En}(Q^2)\to-Q^2R^2/6.$ The quantity $R^2$ is well-measured\cite{nrm} as
$R^2=-0.113 \pm 0.005 \;$fm$^2$. The Galster 
parameterization\cite{galster}
 has been used
to represent the data for $Q^2<0.7\; {\rm GeV}^2$.

Our proton respects charge symmetry, the interchange of $u$ and $d$ quarks,
so  it  contains a prediction for neutron form factors.
This is shown in Fig.~2.
\begin{figure}
\unitlength1cm
\begin{picture}(10,8)(0,-8)
\includegraphics{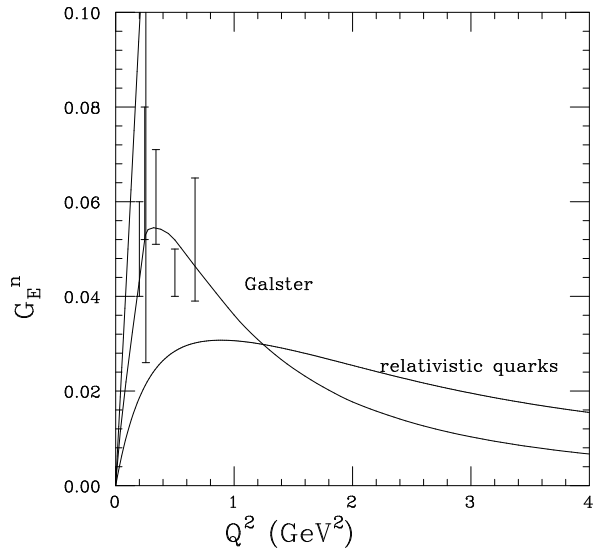}
\end{picture}
\label{fig:2}
\caption{Calculation of $G_E^n.$
  The data
  are from  Ref.~$[13]$, with more expected soon$[14]$. }
\end{figure}
The resulting curve labeled relativistic quarks is both large and small. It is
very small at low values of $Q^2$. Its slope at $Q^2=0$ is too small by a factor
of five, if one compares with the straight line. But at larger values of $Q^2$
the prediction is relatively large.

Our model gives $R^2_{\rm model}=-0.025$ fm$^2$, about five times
smaller than the data. The small value can be understood in terms of $F_{1,2}$.
Taking the  definition (\ref{sachsdefs}) for small values of $Q^2$
gives
\bea-Q^2R^2/6=-Q^2R_1^2/6 -\kappa_n Q^2/4M^2=-Q^2R_1^2/6 -Q^2R_F^2/6,\eea
where the Foldy contribution, 
$R_F^2=6 \kappa_n/4M^2=-0.111\; {\rm fm}^2$, is  in good agreement with
the experimental data. That  a point
particle with a magnetic moment can explain the charge radius has led 
some  to state that $G_E$ is  not a measure
of  the structure of the neutron. However,
one must include the $Q^2$ dependence of $F_1$ which gives $R_1^2$. In our
model $R_1^2=+0.086\; {\rm fm}^2$ which nearly cancels the effects of $R_F^2$.
Isgur\cite{Isgur:1998er} showed that
this cancellation is a natural consequence of including the relativistic
effects of the lower components of the Dirac spinors. Thus our relativistic
effects are standard. We need another source of $R^2$. This is  the
pion cloud.
\section{Pion Cloud  and the Light Front Cloudy Bag Model}
The effects of chiral symmetry require that
sometimes a physical nucleon can be a  bare nucleon immersed in   a pion
cloud. An incident photon can interact electromagnetically with
a bar nucleon, a pion in flight or with a nucleon while a pion is present.
These effects were included in the cloudy bag model\cite{cbm}, and are especially
pronounced for the neutron. Sometimes the neutron can be a proton plus a
negatively charged pion. The tail of the pion distribution extends far out into
space (see Figs. 10 and 11) of Ref.\cite{cbm}, so that the square of the charge
radius is negative.

It is necessary to modernize the cloudy bag model, so as to make it
relativistic. This involves using photon-nucleon form factors from our model,
using a relativistic $\pi$-nucleon form factor, and treating the pionic
contributions relativistically by doing  a light front calculation. 
We define  the resulting model
as the light-front cloudy bag model LFCBM. 

The calculation is implemented 
 by evaluating the relevant Feynman diagrams  of
Fig.~\ref{fig:diagrams} by integrating over
$k^-$ analytically ($k^\mu$ is the momentum of the emitted 
virtual pion) and the other three components numerically;
 see Ref.~\cite{Miller:2002ig}.
 Thus the Feynman graphs, Fig.~\ref{fig:diagrams}, are represented by a single
$\tau$-ordered diagram. The  use of $J^+$ and the
Yan identity\cite{Chang:qi} $S_F(p)=\sum_s
u(p,s)\overline{u}(p,s)/(p^2-m^2+i\epsilon)+\gamma^+/2p^+$
allows one see that the nucleon current operators appearing in
Fig.\ref{fig:diagrams}b
act between on-mass-shell spinors.
\begin{figure}
\unitlength.9cm
\begin{picture}(5,5)(0.9,7.0)
\includegraphics{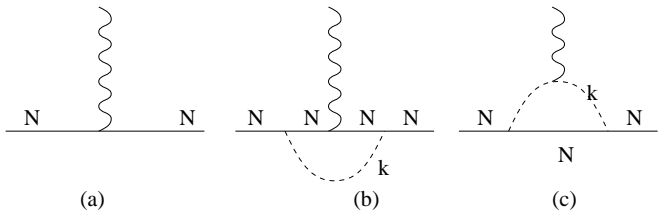}
\end{picture}
\caption{\label{fig:diagrams}Diagrams. The dashed line represents the virtual pion
of  momentum $k^\mu$.}
\end{figure}

 There are four model parameters:$m,\beta,\gamma,\Lambda_{\pi N}$,
with $\Lambda_{\pi N}$ representing
 the cut-off in the pion-nucleon form factor. Including the effects of the 
pion cloud gives contributions to the magnetic 
moments of the proton and neutron,
 so it is necessary to re-fit the parameters.
A sample of the values of the  parameters is given in  Table~1.

\begin{table}
  \centering
  \caption{Different parameter sets, units in terms of  ${\rm fm}$  }
  \vspace{0.1cm}
  \begin{tabular}{|l|rrrrrrr|}\hline
 {\em Set(legend)} & 
 $m$ & $\beta$ & $\Lambda$ & $\gamma$ &-$R^2_n$& $-\mu_n$& $\mu_p$\\ 
\hline
      1 solid & 1.8 &3.65 & 3.1 &4.1&0.111&1.73&2.88\\
      2 dot-dash & 1.7 &3.4& 3.1 &3.9&  0.110&1.79&   2.95\\
      3 dash &1.7&  2.65 & 3.1 & 3.7&   0.109&    1.79          &     2.95\\   
       \hline
      \end{tabular}
      \end{table}

 The result is termed the light front cloudy bag model\cite{Miller:2002ig}
 (even though there is
no bag), 
 and results are shown in Fig.~4. We see that the pion
cloud effects are important for small values of $Q^2$ and, when combined with
those of the relativistic quarks coming from the bare nucleon, leads to a
good description of the low $Q^2$ data. 
The total value of $G_E$ is substantial
for large values of $Q^2$.

\begin{figure}
\unitlength1cm
\begin{picture}(10,8)(0,-8.5)
\includegraphics{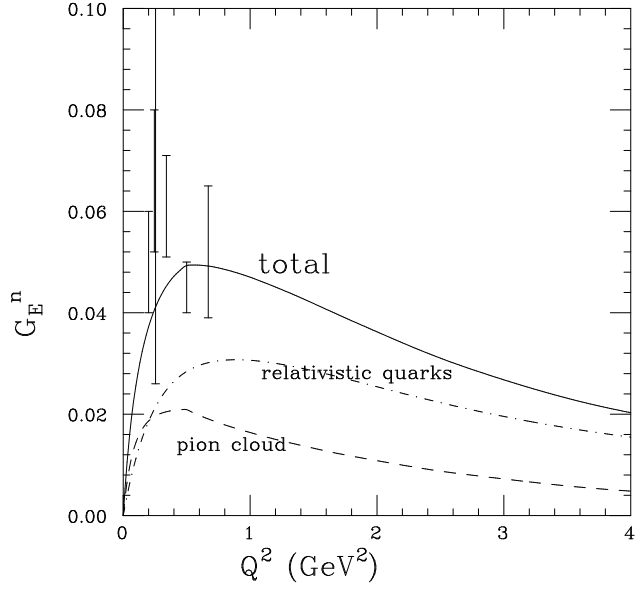}
\end{picture}
\label{fig:ratio}
\caption{Light front cloudy bag model LFCBM Calculation of $G_E^n$[3].}
\end{figure}

One might ask how the effects of the pion cloud influence the results shown in
Fig.~1. They do not change the picture very much. The pion cloud effects
influence the low momentum transfer properties, but vanish at large momentum
transfer. Fig.~1 shows $QF_2/F_1$ which vanishes at $Q=0$, so the influence of
pion cloud effects is hidden.

We have shown one ratio and one form factor. But 
there are four nucleon electromagnetic form factors. The 
remaining two are shown in Figs.~5 and 6.

\begin{figure}
\unitlength1cm
\begin{picture}(10,8)(-1.0,2.5)
\includegraphics{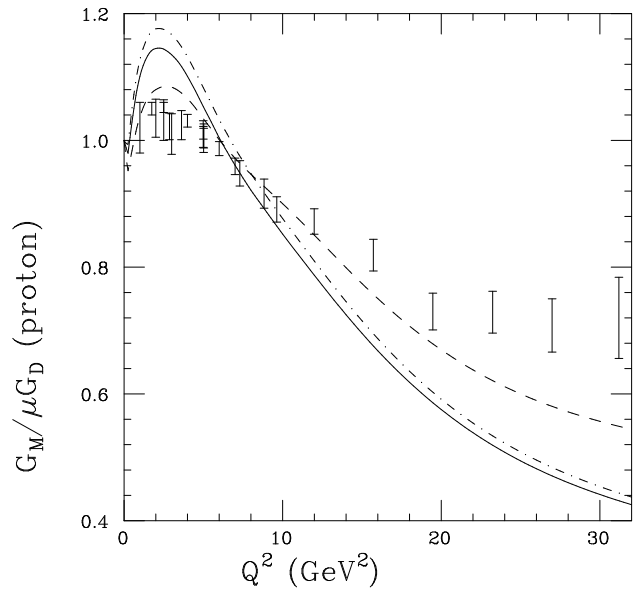}
\end{picture}
\label{fig:gmp}
\caption{Light front cloudy bag model LFCBM Calculation of $G_M^p$[3].}
\end{figure}

\begin{figure}
\unitlength1cm
\begin{picture}(10,8)(1.0,-9.5)
\includegraphics{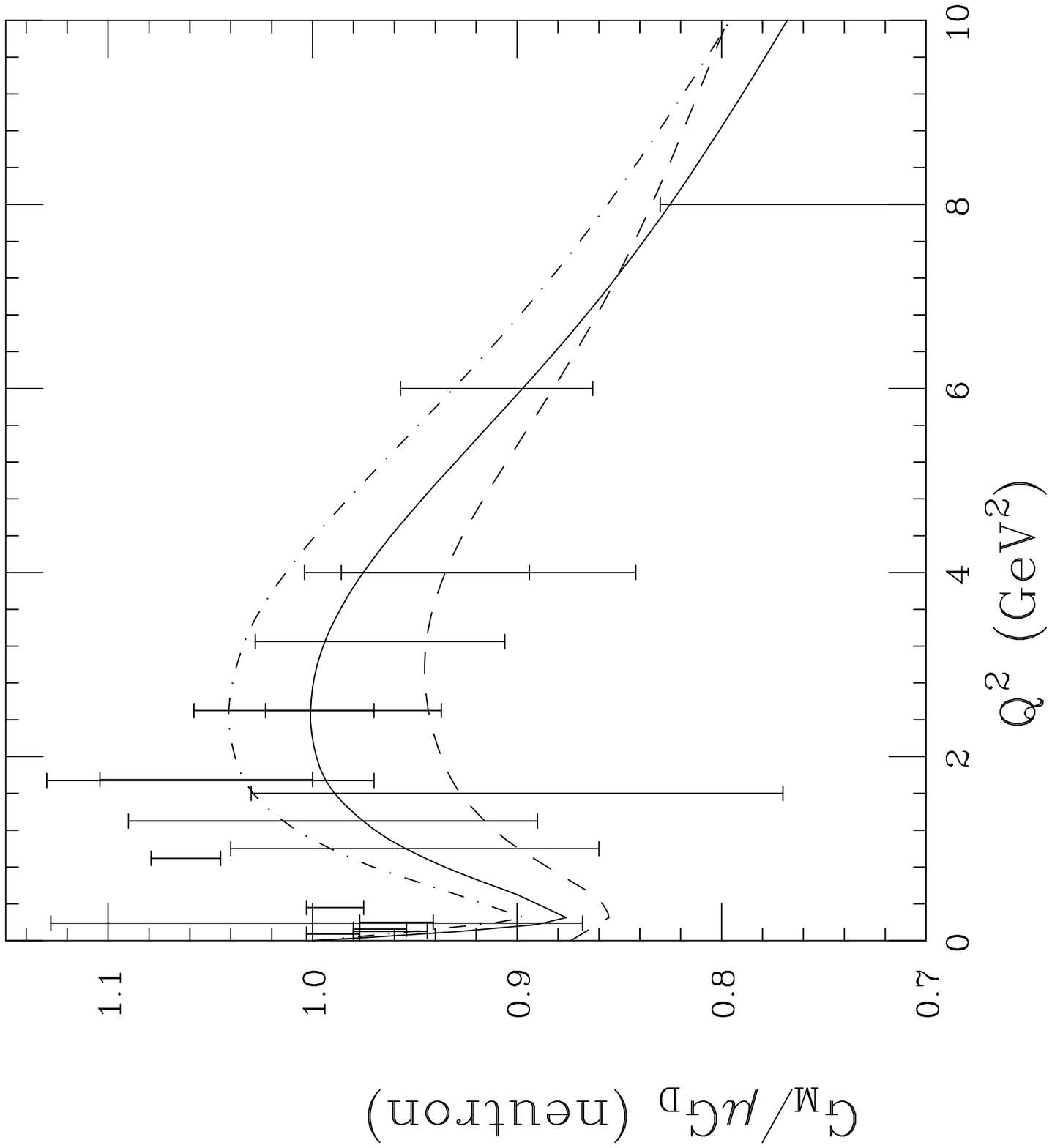}
\end{picture}
\label{fig:gmn}
\caption{Light front cloudy bag model LFCBM Calculation of $G_M^n$[3].}
\end{figure}

Speaking in a general fashion, I can 
say that  good descriptions of $G_M^{p,n}$ are obtained. If one looks closely
there are  disagreements with the data. 
One sees, at very large values of $Q^2$, 
that our value of $G_M^p$ is too small, and this allows room for the effects
of perturbative QCD \cite{BL80}.
 that have been neglected. The wiggles at low values of $Q^2$
are also worthy of comment. These arise 
because the effects 
of the pion field fall off faster with increasing $Q^2$ than the
dipole form used for comparison. Our result for $G_M^n$ shows too deep a dip at
low 
$Q^2$ in comparison with recent data\cite{newgmn}. 

The axial form factor, $G_A(Q^2)$ is not calculated here. Schlumpf's  model
 obtained an excellent reproduction of  existing data\cite{Schlumpf:ce},
 and our parameters are similar to
  his. The lowest-order effect of the pion cloud
 vanishes,
 so the principal difference between $G_A$ of our model 
and that of Ref.~\cite{Schlumpf:ce} is that our quark masses have  larger 
values. This
 increases the computed value of $g_A$ for a bare
 nucleon by about $10-15\%$.
 But this is opposed by the need to multiply the
 bare nucleon result by  the 
 renormalization  factor $Z$ of about $ 0.85-0.9$. Thus our results for 
$G_A(Q^2)$ should
  be  similar  to those of \cite{Schlumpf:ce}.

\section{Summary}
These calculations show  that 
the combination of 
\poinc invariance 
and pion cloud effects,  is sufficient
to describe the existing experimental data up to about $Q^2=20\;{\rm GeV}^2$.
This is somewhat surprising as the model keeps only two  necessary
effects. 
 Configuration  mixing of quark\cite{Cardarelli:2000tk}, 
 the variation of the quark mass with $Q^2$\cite{qvm}, 
exchange currents\cite{weber} and an intermediate $\Delta$\cite{cbm}
 have not been included.  These effects either
  have modest 
 influence,   are incorporated implicitly through the choice
 of parameters, or will help to remove any remaining differences with 
experiment.

\section*{Acknowledgements}
I thank the USDOE for partial support of this work.

\end{document}